\documentclass[twocolumn,preprintnumbers,amsmath,amssymb]{revtex4}

\usepackage{graphicx}
\usepackage{dcolumn}
\usepackage{bm}
\usepackage{epstopdf}

\begin{document}


\title{Coulomb-driven relativistic electron beam compression}

\author{Chao Lu$^{1,2}$, Tao Jiang$^{1,2}$, Shengguang Liu$^{1,2}$, Rui Wang$^{1,2}$, Lingrong Zhao$^{1,2}$, Pengfei Zhu$^{1,2}$, Dao Xiang$^{1,2,3,*}$ and Jie Zhang$^{1,2}$}
\affiliation{%
$^1$ Key Laboratory for Laser Plasmas (Ministry of Education), School of Physics and Astronomy, Shanghai Jiao Tong University, Shanghai 200240, China \\
$^2$  Collaborative Innovation Center of IFSA (CICIFSA), Shanghai Jiao Tong University, Shanghai 200240, China \\
$^3$  Tsung-Dao Lee Institute, Shanghai Jiao Tong University, Shanghai 200240, China \\
}
\date{\today}

\begin{abstract}
Coulomb interaction between charged particles is a well-known phenomenon in many areas of researches. In general the Coulomb repulsion force broadens the pulse width of an electron bunch and limits the temporal resolution of many scientific facilities such as ultrafast electron diffraction and x-ray free-electron lasers. Here we demonstrate a scheme that actually makes use of Coulomb force to compress a relativistic electron beam. Furthermore, we show that the Coulomb-driven bunch compression process does not introduce additional timing jitter, which is in sharp contrast to the conventional radio-frequency buncher technique. Our work not only leads to enhanced temporal resolution in electron beam based ultrafast instruments that may provide new opportunities in probing material systems far from equilibrium, but also opens a promising direction for advanced beam manipulation through self-field interactions.  
\end{abstract}

\maketitle

Ultrafast electron diffraction (UED) in which a fast-evolving process is excited by a pump laser and then probed by a delayed electron pulse has been widely used for watching atoms in motion during structural changes of which the characteristic timescale extends from a few femtoseconds to a few picoseconds \cite{UED1, UED2}. With the advent of ultrashort lasers, the temporal resolution in UED applications now depends primarily on the pulse width and timing jitter of the electron beam. In x-ray free-electron lasers (FELs \cite{FLASH, LCLS, SACLA, FERMI, PAL}), because the x-rays are produced by sending a high-brightness relativistic electron beam through a series of magnetic undulators \cite{PRABreview, NPreview, RMPreview}, the temporal resolution in laser-pump x-ray-probe experiment is also largely dependent on the electron beam pulse width and timing jitter. Therefore, one of the long standing goals in both FEL and UED communities is to generate a high-brightness electron beam with short pulse width and small timing jitter.

According to Coulomb's law, the magnitude of the electric force is directly proportional to the charge and inversely proportional to the square of the distance. So when millions of electrons are produced with a femtosecond laser pulse through photoemission, they tend to repel each other. The Coulomb force in general accelerates the electrons in the front part of the bunch while decelerates those in the back, leading to pulse broadening \cite{JAP}. Fortunately, the elongated electron beam can be compressed with an radio-frequency(rf) buncher \cite{CPRL, CUCLA}. In this method, the electron beam is first sent through an rf buncher cavity where the bunch head is decelerated to lower energy while the bunch tail is accelerated to higher energy; after passing through a dispersive element (typically a drift for keV and MeV electrons or a magnetic chicane for ultra-relativistic electrons (see, e.g. \cite{CSR})), the bunch tail may catch up with the bunch head, leading to bunch compression. While this method has been widely used in UED and FEL community, it is also realized that the phase jitter in the rf cavity leads to considerable timing jitter which limits the temporal resolution achievable \cite{RFC1, RFC2}. The physics behind this effect is rather simple, e.g. in practical cases the bunch centroid does not always pass through the cavity at a constant phase, and the variation of beam energy from this phase jitter is translated to the variation of time-of-flight after passing through the dispersive element. 

    \begin{figure*}[t]
    \includegraphics[width = 0.9\textwidth]{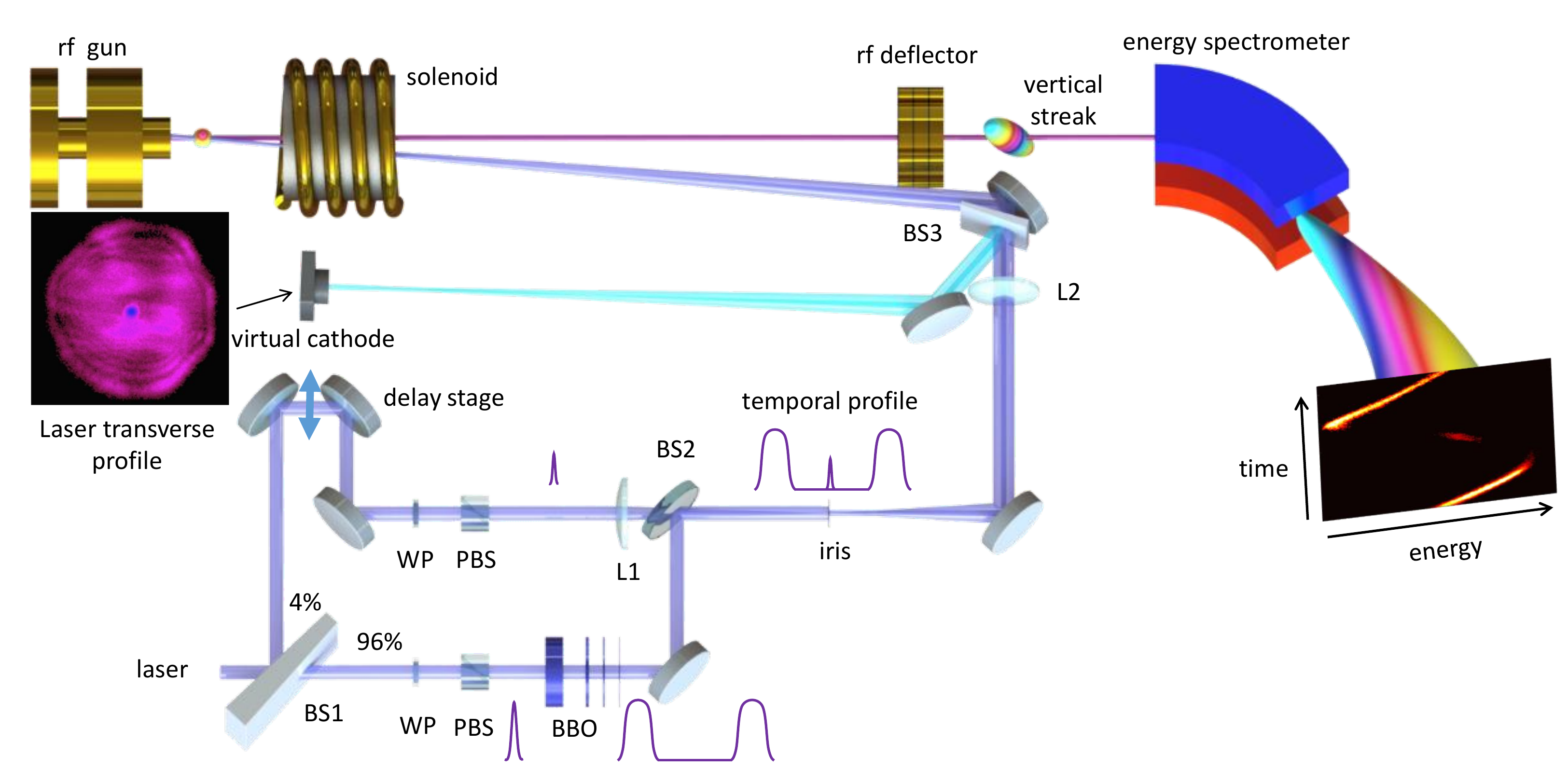}
            \caption{Coulomb-driven relativistic electron beam compression experiment set-up. The laser is first split into two pulses with a beam splitter (BS1) and then combined with a second beam splitter (BS2). The high-energy pulse after pulse shaping is used to produce the \textit{drive beam} and the other low-energy pulse is used to produce the \textit{target beam}. A solenoid lens is used to control the transverse size of the electron beam during beam propagation and the electron beam longitudinal phase space is measured with an rf deflector and an energy spectrometer.  
    \label{Fig.1}}
    \end{figure*}

It should be noted that Coulomb force is not always detrimental. Coulomb force has been used before to suppress beam shot noise at optical wavelength \cite{SNNP, SNPRL}, as well as to enhance the electron beam density modulation at THz wavelength \cite{UCLAPRL, THUPRL}. In this Letter we demonstrate a scheme that makes use of Coulomb force to compress a relativistic electron beam without introducing additional timing jitter. The set-up is shown in Fig.~1. The laser is first split into two pulses with a beam splitter (BS1) so that the energy, pulse width, transverse size and timing of each pulse can be independently adjusted. The main pulse (consisting of 96\% of the energy) is used to produce the \textit{drive beam} and the remaining part is used to produce the \textit{target beam} in the s-band (2856 MHz) photocathode rf gun. After BS1, the main pulse is sent through four $\alpha-$BBO birefringent crystals with the first one (7.5 mm thick) used to produce a double pulse separated by 6.4 ps and the rest three of them (1.4 mm, 0.7 mm and 0.35 mm thick) for shaping the laser into flat-top distribution (2.4 ps FWHM). This laser shaping technique is based on the group velocity mismatch of the ordinary and extraordinary rays in the birefringent crystals and has been widely used for tailoring electron beam distribution for various applications \cite{UCLAPRL, THUPRL, LS1, LS2, ESR}. The energy of each pulse is adjusted with a $\lambda/2$ wave plate (WP) and a polarizing beam splitter (PBS). The high energy pulse and low energy pulse are combined with a second beam splitter (BS2) with 50\%-50\% efficiency. An iris with 1.5 mm diameter is used to produce a uniform distribution for the high energy laser pulse while the low energy laser pulse is focused to 0.2 mm (diameter) at the iris with a lens (L1). The laser distribution at the iris plane is further relayed to the cathode by a lens (L2). The laser transverse profile is measured with a UV camera located at the virtual cathode position where the measured distribution is the same as that on the cathode (the central blue spot in Fig.~1 is the distribution of the low energy pulse). The pulse width of the low energy laser is about 150 fs (rms), limited by the dispersion of the lens, polarizer, and beam splitters.

The relative timing between the drive beam and target beam can be controlled with a delay stage. In this experiment, the target beam is put in the center of the two drive beams for optimal compression. Intuitively the compression may be understood as follows: with the repulsion force from the front and back drive beams, the target beam will stay in the middle with its size shrunk. The physics behind this scheme is illustrated in Fig.~2 which shows the longitudinal component of the Coulomb field of the drive beams. The drive beam is assumed to be uniform in transverse direction with a diameter of 1.5 mm. In longitudinal direction it has a FWHM duration of 2.4 ps and Gaussian ramping (150 fs rms) at the head and tail. The charge is assumed to be 4.5 pC for each of the drive beam. 

    \begin{figure}[b]
    \includegraphics[width = 0.45\textwidth]{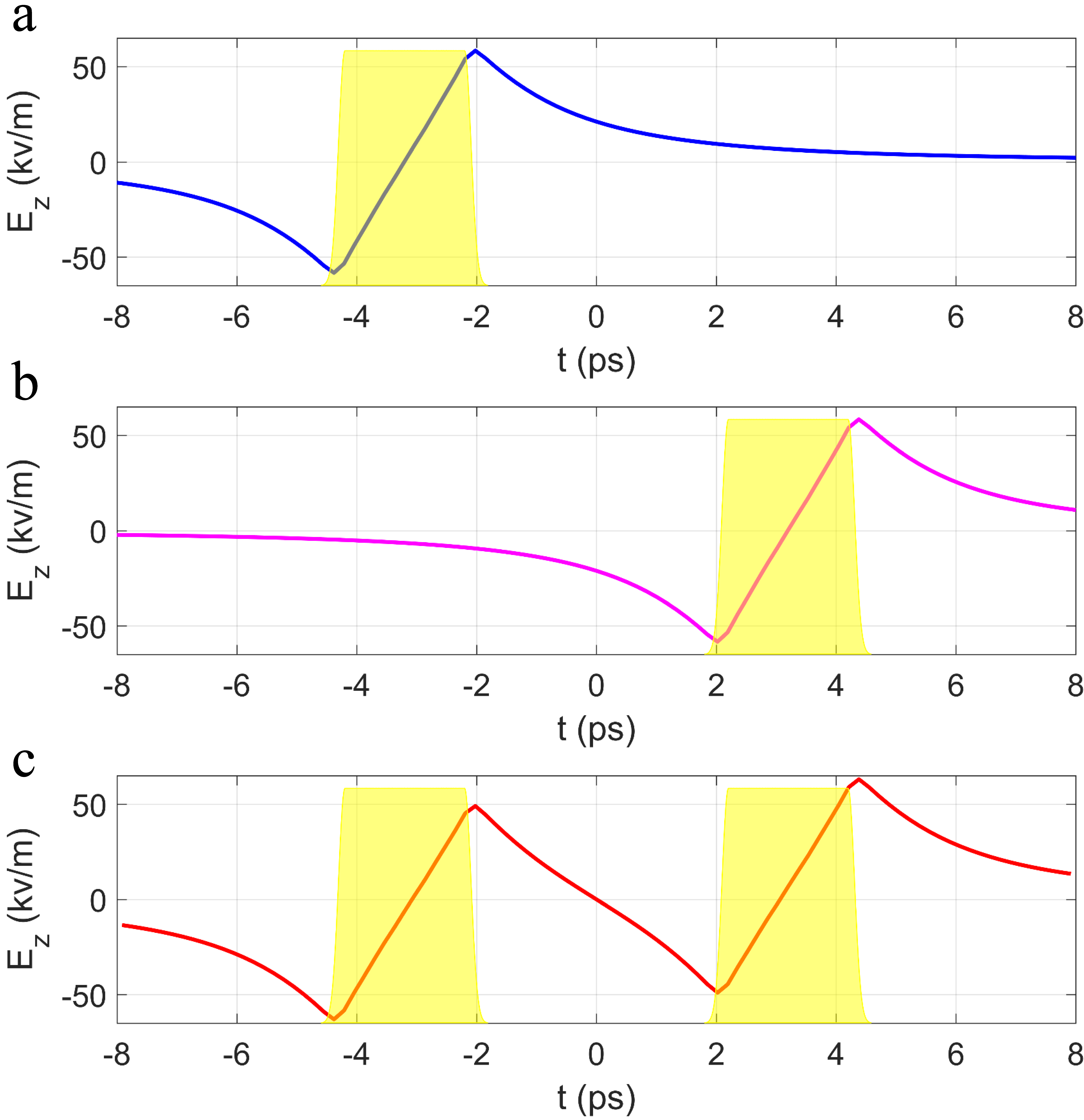}
            \caption{Current distribution (shaded yellow area, bunch head to the right) and the corresponding longitudinal Coulomb field (solid line) for the back drive beam (a), for the front drive beam (b), and when both drive beams are present (c).
    \label{Fig.2}}
    \end{figure}

From Fig.~2 one can see that the longitudinal Coulomb field differs significantly in the region inside the bunch from that outside the bunch. Within the bunch (the shaded area), the longitudinal electric field produces positive energy chirp (correlation between an electron's energy and its longitudinal position; positive chirp meaning bunch head having higher energy than bunch tail) in the beam phase space, which accounts for the well-known bunch lengthening effect. Outside the bunch, a negative chirp (albeit with considerable nonlinearity) will be imprinted, allowing for beam compression in a drift. Specifically, if the target beam is located in the front of the drive beam, it will be pushed further forward by the drive beam and the electrons in the tail of the target beam that are closer to the drive beam will be accelerated more due to the stronger Coulomb force, leading to negative chirp in the target beam phase space. Alternatively, if the target beam is located in the back of the drive beam, it will be pushed further back and the electrons in the head of the target beam will be decelerated more, leading to again negative chirp in the target beam phase space.

    \begin{figure}[b]
    \includegraphics[width = 0.49\textwidth]{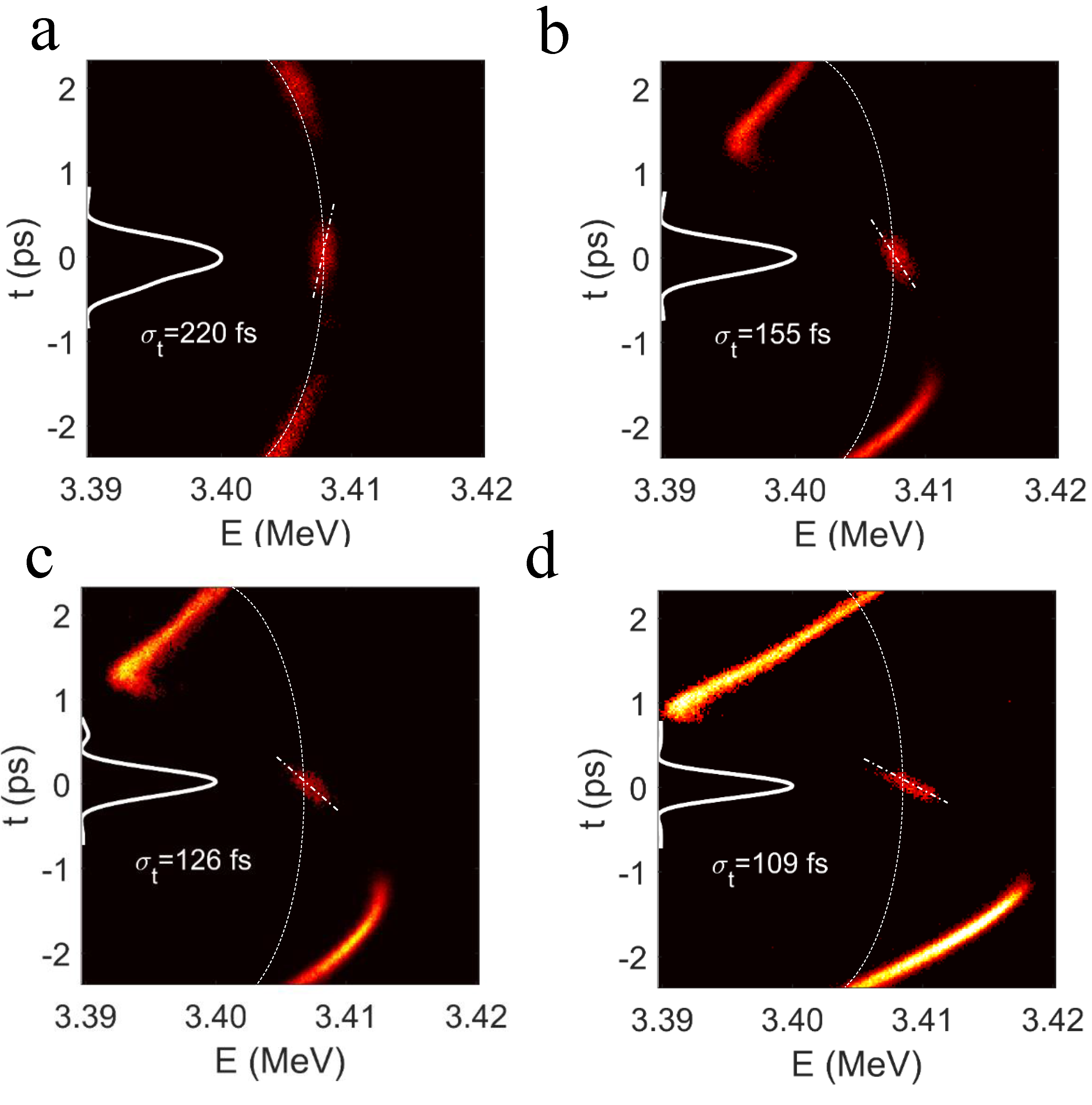}
            \caption{Beam longitudinal phase space (bunch head to the up) as the total charge of the drive beams is gradually increased from 0.2 pC (a) to 3.2 pC (b), 4.5 pC (c) and 6.7 pC (d). The top and bottom distributions show the front drive beam and back drive beam, and the target beam is in the middle. The corresponding electron beam temporal profiles are shown with solid white lines and the rms bunch lengths are also given. Note, due to limited field of view of the screen, only part of the drive beam distribution is measured.  
    \label{Fig.3}}
    \end{figure}

A more optimal scenario is to put the target beam in the middle of two drive beams. As shown in Fig.~2c, the nonlinearity of the negative chirp nearly cancels out for the region in the middle of the drive beams, leading to a linear energy chirp in the target beam phase space, ideal for bunch compression. Furthermore, by symmetry the centroid of the target beam sees zero field which is similar to sending the target beam through an rf cavity at zero-crossing phase. Because the two drive beams are produced by the same laser pulse, the variations in charge, size, length etc. will be the same for both beams. This rigorously locks the target beam to the zero-crossing phase, crucial for avoiding timing jitter. Also the strength of the imprinted chirp is doubled as compared to the single drive beam case, which relaxes the required charge in the drive beam.

    \begin{figure}[b]
    \includegraphics[width = 0.49\textwidth]{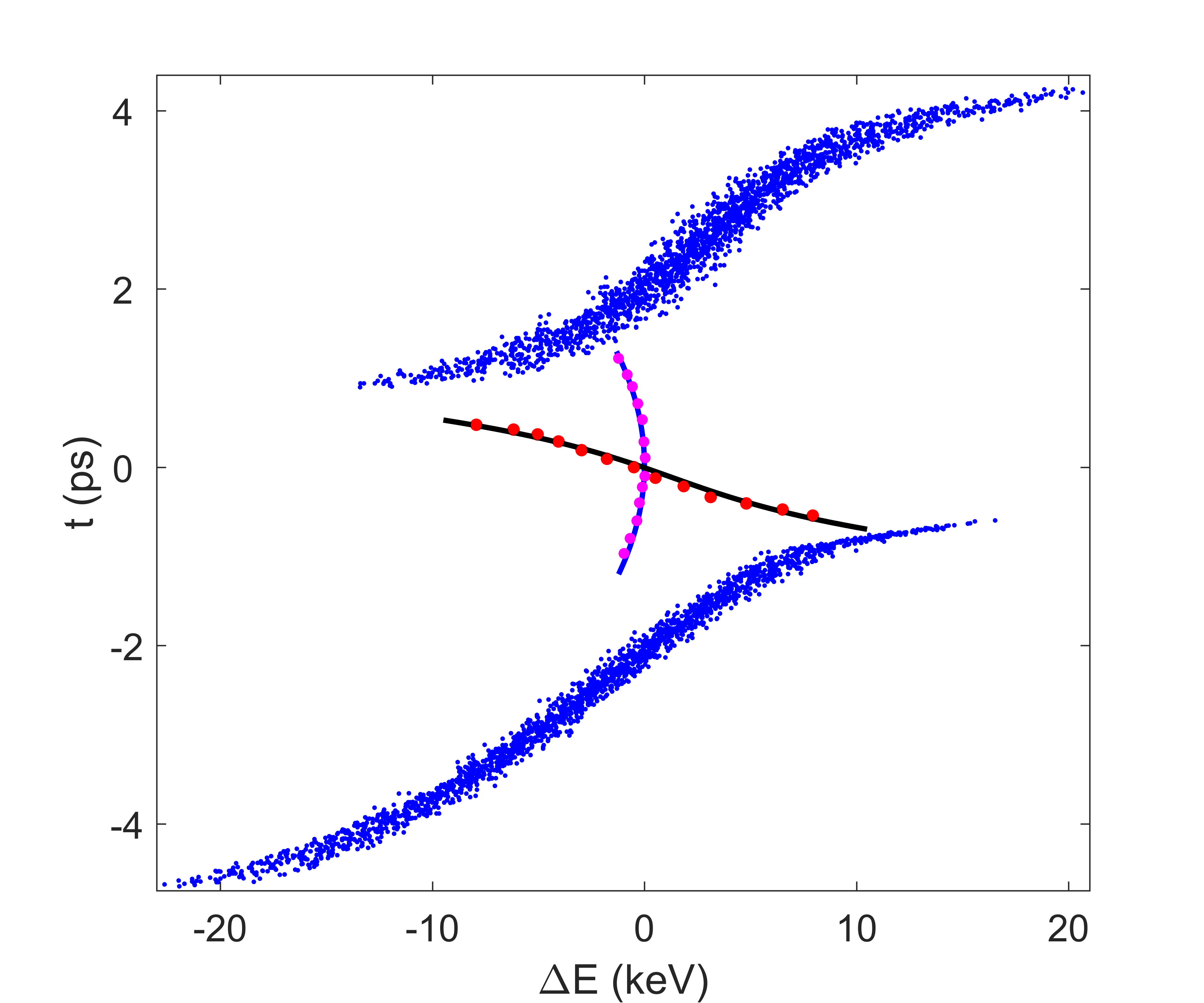}
            \caption{Measured final energy and time of the target beam centroid with (red dot) and without (magenta dot) the drive beams. The simulation results are shown with black (with drive beams) and blue (without the drive beams) lines. The simulated longitudinal phase spaces for the drive beams is shown with blue dots. 
    \label{Fig.4}}
    \end{figure}

The longitudinal phase space for the three electron bunches are measured with a c-band (5712 MHz) rf deflector and an energy spectrometer. When the charge of the drive beam is low (Fig.~3a), its longitudinal phase space just follows the curvature of the sinusoidal rf field (dashed curved line). For the target beam, while its charge is lower than the drive beam, the charge density is higher due to the shorter pulse width and smaller laser size, which leads to slightly positive chirp (illustrated with dashed straight line) in the phase space (Fig.~3a). As we gradually increase the charge of the drive beams (the charge for the target beam, about 50 fC, is constant throughout the experiment), the Coulomb force of the drive beams starts to play a role. As shown in Fig.~3b, Fig.~3c and Fig.~3d, the longitudinal phase space of the drive beams with high charge is dominated by positive energy chirp accounting for bunch lengthening, while the energy chirp of the target beam is reversed to negative enabling bunch compression. The target beam is compressed by about a factor of 2 when the charge of the drive beams is increased to about 6.7 pC.

    \begin{figure*}[t]
    \includegraphics[width = 0.8\textwidth]{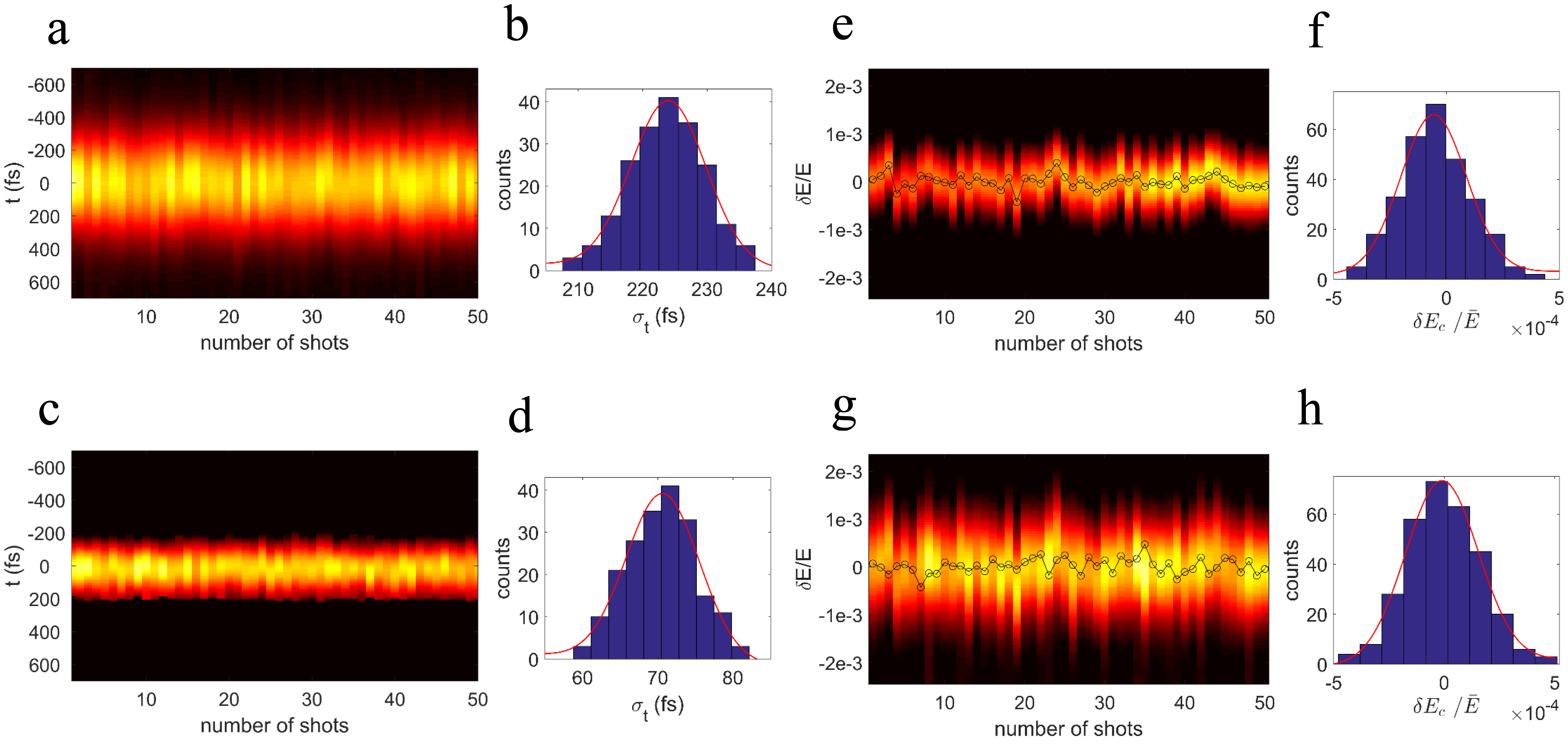}
            \caption{Measured pulse width and energy distribution for the target beam with 50 fC charge. 50 consecutive single-shot measurements of the target beam longitudinal profile without (a) and with (c) the drive beams, and the rms pulse width statistics collected over 200 consecutive shots without (b) and with (d) the drive beams. 50 consecutive single-shot measurements of the target beam energy distribution without (e) and with (g) the drive beams. The centroid beam energy for each individual shot (white circles) is used to evaluate the energy jitter without (f) and with (h) the drive beams.
    \label{Fig.5}}
    \end{figure*}

It should be pointed out that in this experiment the dispersion of the drift is smaller than the optimal value that yields full compression. Analysis shows that the longitudinal phase space of the target beam will continue rotating counter clock-wise if it is further sent through a drift and accordingly reach a much shorter bunch length. It is also worth mentioning that in order to use the target beam for pump-probe applications, it needs to be separated from the drive beams. One straightforward method is to send the beams through an rf deflector of which the rf phase is tuned in such a way that the drive beams are deflected while the target beam passes through at zero-crossing phase; Then a downstream collimator may be used to select the target beam. 

To provide more insights into the Coulomb-driven bunch compression process, the target beam is time delayed with respect to the drive beams (total charge is 9 pC) and the measured correlation of the energy and time of the target beam centroid is shown in Fig.~4 together with the simulation results with the code GPT \cite{GPT}. Specifically, without the drive beams, the target beam energy variation (magenta dot) just follows the quadratic curvature of the sinusoidal rf field. With the Coulomb force from the drive beams, the target beam energy variation (red dot) shows a quasi-linear distribution enabling bunch compression, which is in good agreement with the calculated field in Fig.~2c. 

Finally we show the repeatability of this scheme in generating stable short electron bunches. In this measurement the charge of the drive beams is increased to about 14 pC to increase the compression factor while still maintaining a considerable distance (about 0.5 ps) between the drive beams so that the target beam can be readily separated from the drive beams. The bunch length and energy stability of the target electron beam for many consecutive shots with and without the drive beams are shown in Fig.~5. From Fig.~5 one can see that the target beam has been stably compressed from about 220 fs (rms) to about 70 fs (rms) and the energy stability (both at about $1.6\times10^{-4}$ level) remains essentially unchanged, indicating that no additional timing jitter is introduced during bunch compression. 

In conclusion, we have demonstrated that the Coulomb repulsion effect, widely considered as the driving force for bunch lengthening, can actually be used to compress a relativistic electron beam. The observation of compression by a factor of 3, although moderate, is sufficient to break the 50 fs temporal resolution barrier in UED if one starts with a shorter UV pulse. It is worth pointing out that our results should be considered as representative rather than fully optimized and much higher compression factor may be achieved with more advanced configurations. While the method is demonstrated with MeV electrons and thus is directly applicable to MeV UED \cite{MUED, UCLA, THU, OSAKA, SJTU, BNL, SLAC, DESY}, we expect that this method may also be applied to keV UED and FELs and will stimulate more advanced concepts for enhancing the performance of electron beam based ultrafast instruments.


This work was supported by the Major State Basic Research Development Program of China (Grants No. 2015CB859700) and by the National Natural Science Foundation of China (Grants No. 11327902 and 11504232). One of the authors (DX) would like to thank the support of grant from the office of Science and Technology, Shanghai Municipal Government (No. 16DZ2260200).\\
* dxiang@sjtu.edu.cn

\pagebreak

\end{document}